\documentclass[conference]{IEEEtran}
\IEEEoverridecommandlockouts

\usepackage{cite}
\usepackage{amsmath}
\usepackage{amsfonts}
\usepackage{amssymb}
\usepackage{booktabs}
\usepackage{multirow}
\usepackage{float}
\usepackage[hidelinks,bookmarks=false]{hyperref}
\usepackage[dvipsnames]{xcolor}
\usepackage{flushend}
\usepackage{graphicx}
\usepackage{url}

\begin{document}

\title{Krum-Inspired Central Teacher Selection and Residual Channel Bottlenecks for Efficient DeepSC
 \thanks{This work is supported by the American University of Beirut
 University Research Board (URB) and Vertically Integrated Projects (VIP) Program.}
}

\author{
\IEEEauthorblockN{Rami Eid, Mostafa Jammoul, Omar Kaaki, Maria Slim, Mariette Awad, Hadi Sarieddeen}
\IEEEauthorblockA{Department of Electrical and Computer Engineering \\
American University of Beirut, Beirut, Lebanon \\
Email: \{rae81, mwj06, ohk06, mas194\}@mail.aub.edu, \{ma162, hs139\}@aub.edu.lb}
}

\maketitle

\begin{abstract}

Deploying transformer-based semantic communication models on edge
devices requires compression that preserves semantic fidelity under
channel variability. We study a compressed deep semantic communication
(DeepSC) student trained by multi-teacher knowledge distillation and
combine two ideas: (i) a residual channel bottleneck that splits the
transmitted representation into a base stream and a residual stream
with unequal power allocation, and (ii) a Krum-inspired, medoid-style
centrality criterion that selects a single central teacher from a
five-model ensemble, applied at the logit and intermediate-feature
levels and combined with a feature-dominant distillation loss. On the
EuroParl benchmark, a two-layer student recovers about $98\%$ of the
four-layer teacher's bilingual evaluation understudy (BLEU)-1 under additive white
Gaussian noise, with $93\%$ of its BLEU-4 and $96\%$ of its
sentence-BERT (SBERT) score, and about $89\%$, $77\%$,
and $87\%$ of the teacher's BLEU-1, BLEU-4, and SBERT, respectively,
under Rayleigh fading, while reducing non-embedding parameters by
$1.33\times$ and single-teacher inference latency by $1.79\times$
($9.0\times$ relative to a five-teacher ensemble used here as an
upper-bound reference, not a deployment baseline). Controlled
ablations over three seeds indicate complementary contributions of
about $+8.4\%$ BLEU-1 ($+9.6\%$ SBERT) from the residual bottleneck and
$+3.0\%$ ($+2.9\%$ SBERT) from centrality-based selection over mean
aggregation. A decoder-mode ablation shows that the base stream alone
recovers about two-thirds of the full BLEU-1 while the residual stream
alone collapses, supporting the role of the residual as a refinement
on top of the base.

\end{abstract}


\section{Introduction}
\label{sec:intro}

Semantic communication (SemCom) systems jointly optimize the transmitter
and receiver to preserve task-relevant meaning over noisy channels rather
than exact symbol reconstruction~\cite{weaver1953}. The deep learning-enabled semantic communication (DeepSC) framework~\cite{xie2021}
instantiated this idea with transformer encoder and decoder architectures~\cite{vaswani2017},
with later extensions to images~\cite{bourtsoulatze2019}, speech~\cite{weng2021},
and multimodal content~\cite{wang2025}. Recent work has further highlighted the importance of robust DeepSC
architectures under channel impairments and adversarial perturbations
\cite{alhaj2026signdeepsc}, as well as their generalization across
contemporary channel models~\cite{ismail2026thz}. A practical bottleneck in deploying these models
on edge devices is model size: large transformer pipelines are expensive
to run, while smaller students struggle to preserve semantic fidelity
under channel noise. This paper studies a compressed DeepSC student
trained by multi-teacher knowledge distillation
(KD)~\cite{hinton2015} and asks two specific questions: how to design a
compact channel bottleneck that degrades gracefully with the channel,
and how to select useful supervision from a heterogeneous teacher
ensemble.

Single-teacher KD in SemCom~\cite{liu2024} operates at the output level and does not explicitly supervise the encoder, channel, and decoder pipeline. FitNets
style intermediate-feature supervision~\cite{romero2015} and
multi-teacher ensemble methods~\cite{wu2022,gou2021,hamdi2025} have been
studied extensively in vision and natural language processing, but a
basic difficulty in the SemCom setting is that channel noise causes
individual teachers to produce outputs of varying relative quality on a
given batch: naive averaging can therefore dilute the supervision
signal. In parallel, to the best of our knowledge, prior DeepSC compression approaches use a
single-path channel bottleneck that allocates identical capacity
regardless of channel quality, leaving no built-in graceful-degradation
mechanism.

We address both issues empirically with three contributions, validated
on EuroParl over additive white Gaussian noise (AWGN) and Rayleigh fading with three training seeds.
First, we introduce a residual channel bottleneck that splits the
transmitted representation into a base stream and a refinement residual,
allocating $70\%$ of transmit power to the base. The design is motivated
by unequal error protection (UEP) and the successive-refinement
principle~\cite{equitz1991,rimoldi1994} and validated empirically; it
improves BLEU-1 by $+8.4\%$ over a single-path bottleneck. Second, we adopt a Krum-inspired, medoid-style centrality criterion for teacher selection, scoring each teacher by the sum of squared distances to all other teachers rather than to its closest neighbours as in the original Krum rule~\cite{blanchard2017}. Although Krum was
originally a Byzantine-robust aggregator, in our seed-diversity teacher
pool it converges to a single central teacher per run rather than
adapting per batch (Sec.~\ref{sec:krum_analysis}); we therefore frame
it as a parameter-free, centrality-based selector that yields $+3.0\%$
BLEU-1 over mean aggregation with much lower seed variance than
confidence-weighted averaging. Third, we provide an empirical analysis
of selection behavior, decoder-mode ablations, qualitative recovered
text, and an efficiency analysis whose primary deployment metric is
the $1.79\times$ single-teacher latency reduction; the $9.0\times$
ensemble figure is reported as an upper-bound reference. A negative
result on explicit SNR conditioning is discussed in
Sec.~\ref{sec:film}.

\section{System Model}
\label{sec:system}

Throughout the paper, non-bold letters ($a, A$), bold lowercase letters
($\mathbf{a}$), and bold uppercase letters ($\mathbf{A}$) denote
scalars, vectors, and matrices, respectively. Calligraphic letters
($\mathcal{A}$) denote sets or operators. Superscripts $S$ and $T_k$
distinguish student and the $k$-th teacher quantities.

\subsection{DeepSC Architecture}

For an input sentence $\mathbf{s}=[s_1,\dots,s_L]$ of $L$ tokens, the
DeepSC~\cite{xie2021} semantic encoder produces token embeddings
$\mathbf{p}=f_{\mathrm{enc}}(\mathbf{s})\in\mathbb{R}^{L\times d}$ via
multi-head self-attention, where $d$ is the embedding dimension. A
channel encoder maps these to transmitted symbols
$\mathbf{x}=\mathrm{PN}(f_{\mathrm{ch}}(\mathbf{p}))
\in\mathbb{R}^{L\times d_c}$, where $d_c$ is the per-token
channel-symbol dimension and $\mathrm{PN}(\cdot)$ denotes power
normalization to unit average energy. After transmission, the channel
decoder recovers
$\hat{\mathbf{p}}=g_{\mathrm{ch}}(\mathbf{y})\in\mathbb{R}^{L\times d}$
and the semantic decoder produces output logits
$\hat{\mathbf{s}}=f_{\mathrm{dec}}(\hat{\mathbf{p}})$.

\subsection{Channel Models}

We consider AWGN and flat Rayleigh fading channels:
\begin{align}
 \text{AWGN:}\;&\mathbf{y}=\mathbf{x}+\mathbf{n},\;
 \mathbf{n}\sim\mathcal{N}(\mathbf{0},\sigma^{2}\mathbf{I}),
 \label{eq:awgn}\\[2pt]
 \text{Rayleigh:}\;&\mathbf{y}_{c}=h\,\mathbf{x}_{c}+\mathbf{n}_{c},\;
 h\sim\mathcal{CN}(0,1),
 \label{eq:rayleigh}
\end{align}
where $\mathrm{SNR}_{\mathrm{dB}}$ denotes the signal-to-noise ratio (SNR) in decibels, $\sigma^{2}=10^{-\mathrm{SNR}_{\mathrm{dB}}/10}$, $\mathbf{x}_c$
and $\mathbf{y}_c$ are the complex baseband transmit and receive
sequences, $h\sim\mathcal{CN}(0,1)$ is a unit-variance complex normal
flat-fading coefficient, and
$\mathbf{n}_c\sim\mathcal{CN}(\mathbf{0},\sigma^{2}\mathbf{I})$ is
complex AWGN. Following the standard idealization in the SemCom
literature, perfect channel state information (CSI) is assumed at the
receiver; the receiver applies one-tap equalization
$\hat{\mathbf{x}}_c = \mathbf{y}_c/h$ before reconstruction. Robustness
to imperfect or estimated CSI is left as future work and is acknowledged
as a limitation in Sec.~\ref{sec:limitations}.

\subsection{Problem Statement}

Given $K$ pre-trained teachers $\{T_1,\dots,T_K\}$ with $N_T$ parameters
each, we seek a student $S$ parameterized by $\boldsymbol{\theta}_S$
with $N_S\!\ll\!N_T$ that minimizes the expected distillation loss
over input sentences and channel conditions
\begin{equation}
 \boldsymbol{\theta}_S^{*} =
 \arg\min_{\boldsymbol{\theta}_S}\;
 \mathbb{E}_{\mathbf{s},c,\sigma}\!\big[
 \mathcal{L}_{\mathrm{KD}}\!\big(
 S(\mathbf{s};\boldsymbol{\theta}_S),\,
 \mathcal{A}(T_{1\!:\!K};\mathbf{s},c,\sigma),\,
 \mathbf{s}
 \big)\big],
 \label{eq:obj}
\end{equation}
where the expectation is over input sentence $\mathbf{s}$, channel type
$c\in\{\text{AWGN},\text{Rayleigh}\}$, and noise standard deviation
$\sigma$ sampled from the training SNR distribution ($[5,10]$\,dB; see
Sec.~\ref{sec:method}). The selection operator $\mathcal{A}(\cdot)$ produces supervision targets from the $K$ teacher representations under the same channel condition as the student; we instantiate $\mathcal{A}$ via a Krum-inspired centrality criterion (Sec.~\ref{sec:kd}). The composite loss $\mathcal{L}_{\mathrm{KD}}$ is
defined in~\eqref{eq:loss}.

\section{Proposed Framework}
\label{sec:method}

\begin{figure*}[t]
 \centering
 \includegraphics[width=0.58\textwidth]{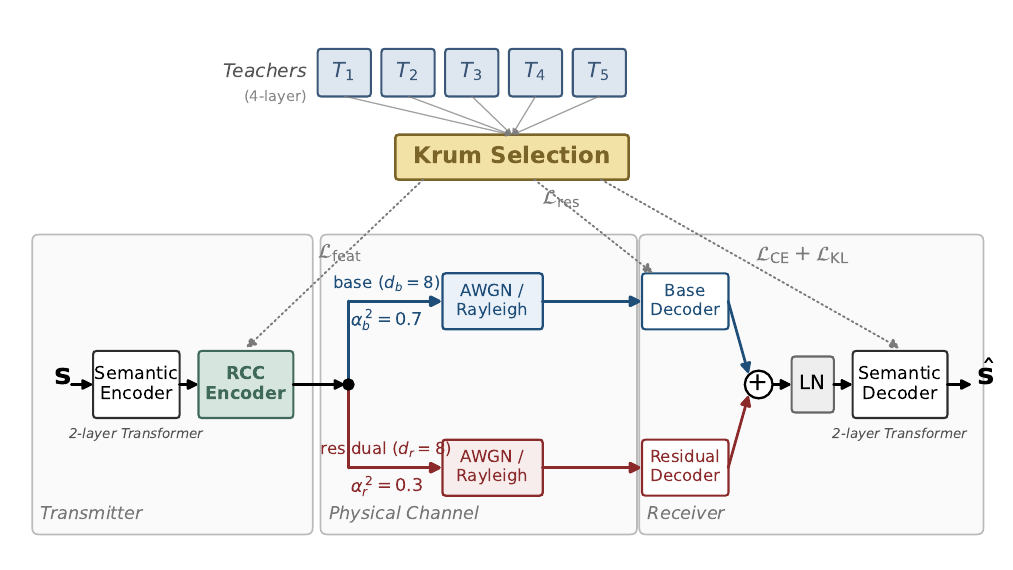}
 \caption{Proposed framework. A five-teacher ensemble is filtered via
 Krum-inspired centrality-based selection to supervise a compressed
 student. The student employs a residual channel bottleneck with a
 base stream ($70\%$ power, coarse semantics) and a residual stream
 ($30\%$ power, correction signal). Dashed lines indicate multi-level
 KD supervision at the encoder output ($\mathcal{L}_{\mathrm{feat}}$),
 channel-decoder output ($\mathcal{L}_{\mathrm{res}}$), and logit
 level ($\mathcal{L}_{\mathrm{CE}},\mathcal{L}_{\mathrm{KL}}$). }
 \label{fig:arch}
\end{figure*}

The proposed framework is illustrated in Fig.~\ref{fig:arch}.

\subsection{Residual Channel Coding}
\label{sec:rcc}

We denote by residual channel coding (RCC) a design in which two independent multilayer perceptrons (MLPs) produce base and residual
codes from the semantic embeddings $\mathbf{p}$:
\begin{align}
 \mathbf{z}_{b}&=\mathrm{PN}(f_{b}(\mathbf{p}))\in\mathbb{R}^{L\times d_b},\\
 \mathbf{z}_{r}&=\mathrm{PN}(f_{r}(\mathbf{p}))\in\mathbb{R}^{L\times d_r},
\end{align}
with $d_b=d_r=8$ and $d_b+d_r=16$ matching the teacher channel-symbol
dimension $d_c$, preserving spectral bandwidth.

The design follows the principle of successive refinement of
information~\cite{equitz1991}: a coarse description should be
sufficient to provide a usable reconstruction floor under impairment,
while a refinement layer can tolerate higher error rates because it acts
only as a correction conditional on the base. This decomposition is also
closely related to layered and scalable source coding in classical
multimedia~\cite{rimoldi1994}. We instantiate this with differentiated
power scaling, also termed unequal error protection (UEP):
\begin{equation}
 \tilde{\mathbf{z}}_b=\alpha_b\mathbf{z}_b,\quad
 \tilde{\mathbf{z}}_r=\alpha_r\mathbf{z}_r,\quad
 \frac{\alpha_b^2}{\alpha_r^2}=\frac{7}{3},\quad
 \alpha_b^2+\alpha_r^2=2,
 \label{eq:uep}
\end{equation}
 where the second constraint enforces unit total transmit power,
allocating $70\%$ of the transmit power to the base stream. The base
stream therefore dominates end-to-end distortion under noisy
conditions, while the residual contributes a refinement that improves
reconstruction when channel conditions allow. The $70/30$ power split
and dimension allocation are motivated by UEP and validated empirically
(Sec.~\ref{sec:rate}). The main results in Table~\ref{tab:main} use a
balanced $d_b\!=\!d_r\!=\!8$ split: equal stream capacity makes the
decoder-mode ablation in Sec.~\ref{sec:ablation} (base-only vs.\
residual-only) attribute any gap purely to UEP and the loss design,
not a dimension imbalance. Sec.~\ref{sec:rate} reports a small
additional gain when the split is shifted toward the base
($d_b\!=\!12$, $d_r\!=\!4$), consistent with the same UEP intuition.

The receiver performs coarse-to-fine reconstruction:
\begin{equation}
 \hat{\mathbf{p}}=
 \mathrm{LN}\!\big(g_b(\hat{\mathbf{z}}_b)+g_r(\hat{\mathbf{z}}_r)\big),
 \label{eq:decode}
\end{equation}
where $g_b,g_r$ are MLP decoders and $\mathrm{LN}$ performs layer
normalization. Both streams share the same channel realization within each
mini-batch to prevent exploitation of realization diversity. The
student backbone uses two Transformer layers (vs.\ four in teachers),
$d_{\mathrm{model}}\!=\!128$, $8$ heads, and feed-forward width
$d_{\mathrm{ff}}\!=\!256$ (vs.\ $512$).

\subsection{Centrality-Based Multi-Teacher Knowledge Distillation}
\label{sec:kd}

Given $K\!=\!5$ teachers producing per-batch representations
$\{\mathbf{z}_k\}_{k=1}^{K}$ at a chosen representation level (logits or
intermediate features), the Krum-inspired centrality criterion returns
the index of the most central teacher:
\begin{equation}
 k^{*}(\mathbf{z}_{1\!:\!K})=
 \arg\min_{k\in[K]}\sum_{j\neq k}\|\mathbf{z}_k-\mathbf{z}_j\|^{2}.
 \label{eq:krum}
\end{equation}
Originally developed for Byzantine-resilient distributed learning, we
re-use this criterion as a medoid-style selector. Although evaluated
per batch, in our seed-diversity teacher pool the same teacher is
chosen at every batch within a run (Sec.~\ref{sec:krum_analysis}); we
therefore frame it as a centrality-based selector rather than a
per-batch robust aggregator. The criterion is evaluated independently
at three representation levels: output logits $\hat{\mathbf{s}}^{T_k}$,
channel-decoder features $\hat{\mathbf{p}}^{T_k}$, and final-decoder
hidden states $\hat{\mathbf{t}}^{T_k}$, with only the selected
teacher's representation used as the target at each level. We do not claim
this is optimal for KD; it is a parameter-free choice that compares
favorably to two natural alternatives: mean aggregation, which weights
all teachers equally regardless of relative quality, and
confidence-weighted averaging (CWA), which is sensitive to teacher
entropy calibration.

The student minimizes a weighted sum of four losses:
\begin{equation}
 \mathcal{L}=
 \lambda_{\mathrm{CE}}\,\mathcal{L}_{\mathrm{CE}}
 +\lambda_{\mathrm{KL}}\,\mathcal{L}_{\mathrm{KL}}
 +\lambda_{\mathrm{feat}}\,\mathcal{L}_{\mathrm{feat}}
 +\lambda_{\mathrm{res}}\,\mathcal{L}_{\mathrm{res}},
 \label{eq:loss}
\end{equation}
with non-negative weights
$(\lambda_{\mathrm{CE}},\lambda_{\mathrm{KL}},\lambda_{\mathrm{feat}},
\lambda_{\mathrm{res}})=(0.25,0.10,0.55,0.10)$
 summing to one, and
softening temperature $\tau\!=\!2$. These weights were selected on a
small validation grid: $\lambda_{\mathrm{feat}}$ was explored over
$\{0.30,0.40,0.55,0.70\}$ with the remaining weights re-balanced
proportionally to preserve the unit-sum constraint, and the chosen
$\lambda_{\mathrm{feat}}=0.55$ yielded the best validation BLEU-1. The
dominance of the intermediate-feature term reflects empirical evidence
that representation alignment, rather than logit matching alone, drives
semantic KD quality in this setup.

The cross-entropy (CE) term $\mathcal{L}_{\mathrm{CE}}$ supervises against ground-truth tokens. The Kullback--Leibler (KL) term matches softened distributions:
\begin{equation}
 \mathcal{L}_{\mathrm{KL}}=\tau^{2}\,
 \mathrm{KL}\Big(
 \mathrm{softmax}\big(\tfrac{\hat{\mathbf{s}}^{T_{k^*}}}{\tau}\big)
 \,\Big\|\,
 \mathrm{softmax}\big(\tfrac{\hat{\mathbf{s}}^S}{\tau}\big)
 \Big),
 \label{eq:kl}
\end{equation}
 where $T_{k^*}$ is the centrality-selected teacher at the logit level. The
feature loss aligns three intermediate representations:
\begin{equation}
 \mathcal{L}_{\mathrm{feat}}=
 w_p\|\mathbf{p}^S\!-\!\mathbf{p}^{T_{k^*}}\|^2
 +w_{\hat p}\|\hat{\mathbf{p}}^S\!-\!\hat{\mathbf{p}}^{T_{k^*}}\|^2
 +w_t\,\mathcal{D}_{\cos}(\hat{\mathbf{t}}^S,\hat{\mathbf{t}}^{T_{k^*}}),
 \label{eq:feat}
\end{equation}
with $(w_p,w_{\hat p},w_t)=(0.2,0.5,0.3)$, emphasizing the
channel-decoder representation, and $\mathcal{D}_{\cos}(\mathbf{u},\mathbf{v})=
1-\langle\mathbf{u},\mathbf{v}\rangle/(\|\mathbf{u}\|\|\mathbf{v}\|)$
denoting cosine distance. Each teacher target in~\eqref{eq:feat} is the
centrality-selected teacher at that respective representation level; with a
slight abuse of notation we use the same symbol $T_{k^*}$ at each
level. The residual loss trains the correction stream:
\begin{equation}
 \mathcal{L}_{\mathrm{res}}=\big\|
 (\hat{\mathbf{p}}-\hat{\mathbf{p}}_{\mathrm{rough}})
 -(\hat{\mathbf{p}}^{T_{k^*}}-\hat{\mathbf{p}}_{\mathrm{rough}}^{\,\mathrm{sg}})
 \big\|^2,
 \label{eq:res}
\end{equation}
where $\mathrm{sg}$ denotes stop-gradient.

\subsection{Training Protocol}

Five teachers (4-layer DeepSC) are pre-trained with different random
seeds on EuroParl~\cite{koehn2005} for $40$ epochs under Rayleigh
fading (Adam, learning rate $\mathrm{lr}=2\!\times\!10^{-4}$). Seed-based diversity
is standard in ensemble KD~\cite{wu2022,gou2021} and suffices here
because seed-induced variation already produces measurable divergence
in representation space under channel noise
(Sec.~\ref{sec:krum_analysis}). We additionally explored stronger
diversity by training fresh teachers across heterogeneous (channel,
SNR) regimes; under a fixed compute budget this degraded student
BLEU-1 by roughly $0.04$--$0.06$ on AWGN at the same student capacity,
indicating that maintaining individual teacher quality is empirically
more important than diversifying their training distributions. Student training uses Adam ($\mathrm{lr}=10^{-4}$, weight decay
$10^{-4}$, batch~$384$, A100 graphics processing unit (GPU), $40$ epochs). Training SNR is
sampled uniformly at random in $[5,10]$\,dB. Mixed-precision training is enabled, and \texttt{torch.compile} is used for execution optimization. The
vocabulary contains $22{,}234$ tokens. Evaluation uses
$\mathrm{SNR}\!\in\!\{0,3,6,9,12,18\}$\,dB with BLEU-1, BLEU-4, and
sentence-BERT (SBERT)~\cite{reimers2019} cosine similarity, providing
both surface-form (n-gram overlap) and meaning-level views of fidelity.

\section{Experimental Results}
\label{sec:results}

\subsection{Baselines and Evaluation Setup}

We compare four configurations. (i) Ensemble: a reference upper bound,
not a deployment baseline, that processes each evaluation batch
through all five 4-layer teachers under the same channel condition
$(c,\sigma)$, then averages their per-token softmax distributions
$\hat{\mathbf{p}}_{\mathrm{ens}}(\cdot)=\frac{1}{K}\sum_{k=1}^{K}
\mathrm{softmax}(\hat{\mathbf{s}}^{T_k})$, with greedy decoding picking
the argmax. The ensemble incurs $5\times$ the cost of a single teacher.
(ii) DeepSC Teacher: a single 4-layer teacher (no compression). (iii)
Proposed: the 2-layer student with RCC and centrality-based KD. (iv)
SingleKD: a 2-layer student distilled from one teacher with logit-only
supervision (lower-bound reference); component-isolating comparisons
appear in Table~\ref{tab:ablation}. Learned configurations are
reported as mean$\,\pm\,$std over three independent training runs;
Ensemble and DeepSC Teacher reuse fixed pre-trained models.

\subsection{Main Results}

Table~\ref{tab:main} presents the primary comparison. Under AWGN, the
proposed student achieves $0.786\pm0.003$ average BLEU-1, recovering
$97.6\%$ of the four-layer teacher ($0.805$) at $1.33\times$ fewer
non-embedding parameters and $1.79\times$ lower single-teacher latency
(Sec.~\ref{sec:eff}); the $9.0\times$ ratio is relative to the five-teacher ensemble. SingleKD reaches only $0.41\pm0.04$ BLEU-1 under
AWGN and $0.29\pm0.05$ under Rayleigh, substantially below both the
proposed system and the teacher; logit-only single-teacher supervision
is thus insufficient to transfer the encoder, channel, and decoder
structure of the teacher to a two-layer student. Under Rayleigh fading
the proposed student reaches $0.603\pm0.023$ BLEU-1; the wider gap
reflects fading severity and is plausibly attributable to the fixed
UEP ratio targeting average channel conditions: under deep fades the
base stream itself suffers corruption that the residual cannot fully
compensate, suggesting adaptive power allocation conditioned on
instantaneous CSI as future work.

\begin{table}[!t]
\centering
\caption{Performance comparison averaged over
 $\mathrm{SNR}\in\{0,3,6,9,12,18\}$\,dB. Proposed and SingleKD show
 mean$\,\pm\,$std over three independent training seeds.}
\label{tab:main}
\setlength{\tabcolsep}{3.0pt}
\begin{tabular}{@{}llccc@{}}
\toprule
\textbf{Channel} & \textbf{Method}
 & \textbf{BLEU-1} & \textbf{BLEU-4} & \textbf{SBERT}\\
\midrule
\multirow{4}{*}{AWGN}
 & Ensemble (5T)
 & 0.889 & 0.750 & 0.818 \\
 & DeepSC Teacher
 & 0.805 & 0.606 & 0.743 \\
 & \textbf{Proposed}
 & \textbf{0.786 {\scriptsize$\pm$0.003}}
 & \textbf{0.564 {\scriptsize$\pm$0.004}}
 & \textbf{0.716 {\scriptsize$\pm$0.004}}\\
 & SingleKD
 & 0.410 {\scriptsize$\pm$0.039}
 & 0.106 {\scriptsize$\pm$0.034}
 & 0.426 {\scriptsize$\pm$0.018}\\
\midrule
\multirow{4}{*}{Rayleigh}
 & Ensemble (5T)
 & 0.823 & 0.673 & 0.774 \\
 & DeepSC Teacher
 & 0.680 & 0.465 & 0.659 \\
 & \textbf{Proposed}
 & \textbf{0.603 {\scriptsize$\pm$0.023}}
 & \textbf{0.356 {\scriptsize$\pm$0.035}}
 & \textbf{0.572 {\scriptsize$\pm$0.020}}\\
 & SingleKD
 & 0.289 {\scriptsize$\pm$0.052}
 & 0.045 {\scriptsize$\pm$0.016}
 & 0.377 {\scriptsize$\pm$0.018}\\
\bottomrule
\end{tabular}
\end{table}

\begin{figure}[!t]
 \centering
 \includegraphics[width=0.74\columnwidth]{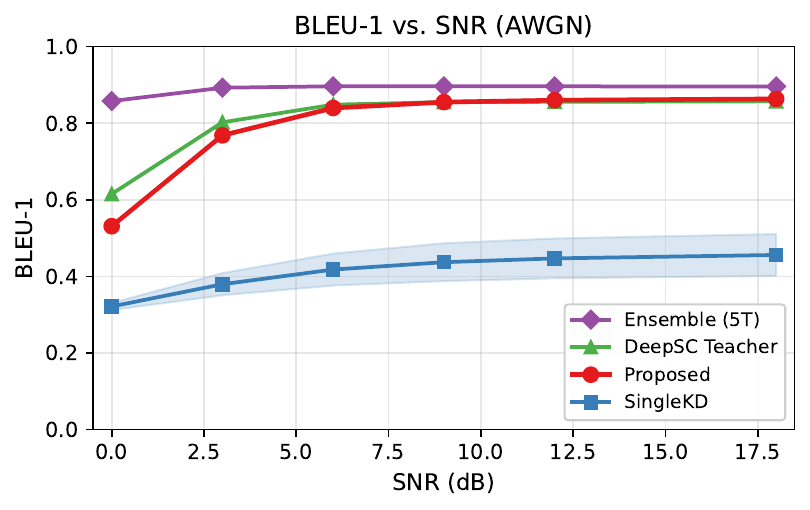}
 \caption{BLEU-1 vs.\ SNR under AWGN. Proposed and SingleKD
 curves show the mean across three seeds; the shaded bands are
 $\pm 1$\,std.}
 \label{fig:snr_awgn}
\end{figure}

\begin{figure}[!t]
 \centering
 \includegraphics[width=0.74\columnwidth]{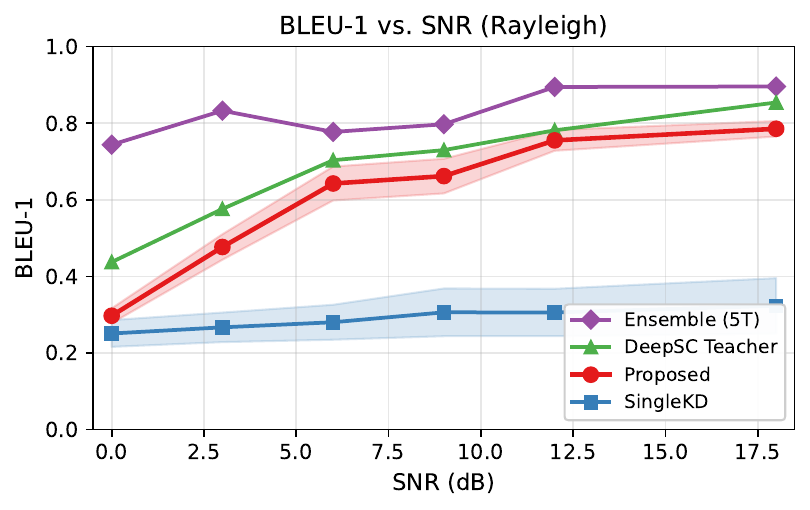}
 \caption{BLEU-1 vs.\ SNR under Rayleigh fading. Proposed and SingleKD
 curves show the mean across three seeds; the shaded bands are
 $\pm 1$\,std.}
 \label{fig:snr_rayleigh}
\end{figure}

Figs.~\ref{fig:snr_awgn} and~\ref{fig:snr_rayleigh} show per-SNR
BLEU-1. Under AWGN, the proposed student nearly matches the teacher at
$\mathrm{SNR}\!\geq\!6$\,dB ($0.840$ vs.\ $0.848$), with the steepest
improvement between $0$ and $6$\,dB. Under Rayleigh it matches the
teacher above $9$\,dB, with the gap widening below $6$\,dB consistent
with the base stream bearing the full channel burden under deep fades.
SingleKD stays in the $0.30$ to $0.40$ BLEU-1 range with little
SNR-driven structure and substantially larger seed variance
($\pm0.04$ to $\pm0.05$ vs.\ $\pm0.003$ for the proposed configuration
on AWGN).

\subsection{Ablation Study}
\label{sec:ablation}

Table~\ref{tab:ablation} isolates each system component under AWGN with
$40$-epoch training and $\mathrm{SNR}\!\in\!\{0,6,12,18\}$\,dB,
reported as mean$\,\pm\,$std over three seeds. The Residual+Mean
configuration uses identical architecture and loss, replacing only
Krum with mean aggregation. Krum gains $+0.022$ BLEU-1 ($+3.0\%$;
$+0.020$ SBERT, $+2.9\%$), larger than the seed std of either
configuration ($0.002$ for Krum, $0.005$ for Mean). On a standard
bottleneck, Krum vs.\ CWA gives $+9.3\%$ BLEU-1 ($+10.2\%$ SBERT), and
the residual bottleneck vs.\ standard with Krum fixed gives $+8.4\%$
($+9.6\%$ SBERT); BLEU-1 and SBERT track closely, mitigating the
concern that the gains are an artifact of unigram overlap. The
Standard+CWA configuration also has roughly $15\times$ the seed std of
the Krum-based rows ($\pm0.033$ vs.\ $\pm0.002$ to $\pm0.006$),
evidence of CWA instability across seeds.

\begin{table}[!t]
\centering
\caption{Ablation under AWGN ($40$ epochs, mean$\,\pm\,$std over $3$
 seeds, averaged over $\mathrm{SNR}\in\{0,6,12,18\}$\,dB). Here, CWA
 denotes confidence-weighted averaging of teacher distributions.}
\label{tab:ablation}
\setlength{\tabcolsep}{3.0pt}
\begin{tabular}{@{}lccc@{}}
\toprule
\textbf{Configuration}
 & \textbf{BLEU-1} & \textbf{BLEU-4} & \textbf{SBERT}\\
\midrule
Residual + Krum (full)
 & \textbf{0.773 {\scriptsize$\pm$0.002}}
 & \textbf{0.549 {\scriptsize$\pm$0.003}}
 & \textbf{0.709 {\scriptsize$\pm$0.004}}\\
Residual + Mean
 & 0.751 {\scriptsize$\pm$0.005}
 & 0.515 {\scriptsize$\pm$0.008}
 & 0.689 {\scriptsize$\pm$0.008}\\
Standard + Krum
 & 0.713 {\scriptsize$\pm$0.006}
 & 0.461 {\scriptsize$\pm$0.008}
 & 0.647 {\scriptsize$\pm$0.006}\\
Standard + CWA
 & 0.653 {\scriptsize$\pm$0.033}
 & 0.371 {\scriptsize$\pm$0.042}
 & 0.587 {\scriptsize$\pm$0.031}\\
\midrule
\multicolumn{4}{@{}l@{}}{\small Isolated gains (BLEU-1, mean of means):}\\
\multicolumn{4}{@{}l@{}}{\small\hspace{1em}
 Krum vs.\ Mean (residual fixed): $+0.022$ ($+3.0\%$)}\\
\multicolumn{4}{@{}l@{}}{\small\hspace{1em}
 Residual vs.\ Standard (Krum fixed): $+0.060$ ($+8.4\%$)}\\
\multicolumn{4}{@{}l@{}}{\small\hspace{1em}
 Krum vs.\ CWA (standard bottleneck): $+0.061$ ($+9.3\%$)}\\
\bottomrule
\end{tabular}
\end{table}

A complementary decoder-mode ablation isolates the contribution of
each stream on a representative trained student (seed~$42$, AWGN, six
SNRs). Decoding from the base stream alone (zeroing
$\hat{\mathbf{z}}_r$ in~\eqref{eq:decode}) recovers BLEU-1~$=0.489$ and
SBERT~$=0.507$, about $64\%$ of the full-decoding BLEU-1 on this seed
($0.760$): a usable coarse reconstruction. Decoding from the residual
alone (zeroing $\hat{\mathbf{z}}_b$) collapses to BLEU-1~$=0.000$ with
SBERT~$=0.21$, near the score of unrelated sentence pairs. The
residual therefore operates strictly as a refinement on top of the
base, matching the UEP and successive-refinement design intent.

\subsection{Selection Behavior}
\label{sec:krum_analysis}

We logged the selected teacher index per batch across all six
evaluation SNRs. In a given run, one teacher is selected with $100\%$
frequency at every SNR. Pairwise distance analysis on a held-out
validation batch explains this: in logit space, the mean $L_2$
distance from the most-central teacher to the remaining four is about
$4.5\%$ smaller than that of the most peripheral teacher, and the
ordering is consistent across batches and SNRs within a run. The
criterion thus behaves as a centrality-based selector that identifies
a single anchor per run rather than a per-batch robust filter; the
empirical claim is centrality, not Byzantine robustness. We further
observed that the same teacher wins at all three representation levels
(logits, channel-decoder features, decoder hidden states); the
multi-level evaluation is retained in case heterogeneous teacher pools
yield level-dependent winners.

To check whether the system depends on a specific teacher, we removed
$T_1$ and re-trained on the remaining four. The criterion converged
on a different central teacher and the student achieved BLEU-1~$=0.800$
on AWGN over six SNRs, comparable to $0.786\pm0.003$ for the full
setup. The selector tracks the centroid of whichever ensemble is
provided rather than a fixed teacher.

\subsection{Qualitative Recovered Text}
\label{sec:qualitative}

Table~\ref{tab:qual} shows two test sentences and their recovered
text under AWGN at three SNR levels. At $18$\,dB the system recovers
the short example verbatim and preserves the clause structure of the
longer example with occasional named-entity slips. At $6$\,dB the
structure is largely retained while content words drift; at $0$\,dB
the output degrades to coarse topical fragments while remaining
locally fluent. The pattern matches the BLEU and SBERT trends in
Figs.~\ref{fig:snr_awgn} and~\ref{fig:snr_rayleigh}.

\begin{table}[!t]
\centering
\caption{Qualitative AWGN recovery on two test sentences.}
\label{tab:qual}
\setlength{\tabcolsep}{3pt}
\renewcommand{\arraystretch}{1.00}
\footnotesize
\begin{tabular}{@{}p{0.13\columnwidth} p{0.79\columnwidth}@{}}
\toprule
\textbf{SNR} & \textbf{Recovered text} \\
\midrule
src & priorities for the budget section iii commission \\
0\,dB & responsibility for the impact green kingdom commission \\
6, 18\,dB & priorities for the budget section iii commission \\
\midrule
src & thank you for the suggestion mr bradbourn we will pass your observations to the quaestors so that they may take the necessary measures \\
0\,dB & you today for the chamber sv we we have have your idea to the fact do that they have take the necessary no so they have \\
6\,dB & thank you for the author mr falconer we will pass your remarks to the honour so that they may take the necessary measures \\
18\,dB & thank you for the original mr falconer we will pass your remarks to the honour so that they may take the necessary measures \\
\bottomrule
\end{tabular}
\end{table}

\subsection{Explicit SNR Conditioning Does Not Improve in This Setup}
\label{sec:film}

A natural question is whether the student benefits from explicit SNR
awareness. We tested two conditioning mechanisms: (i) FiLM-style affine
modulation~\cite{perez2018} of decoded features by an SNR embedding,
and (ii) an adaptive power gate learning $(\alpha_b,\alpha_r)$ as a
function of SNR. Neither improved over the fixed $70/30$ UEP design:
the adaptive gate converged to a near-constant $65.5/34.5$ split
across all SNRs (BLEU-1: $0.780$ vs.\ $0.785$ for the fixed split),
suggesting the UEP-motivated $70/30$ ratio is close to optimal under
our training distribution. A plausible explanation is that the
feature-dominant loss ($\lambda_{\mathrm{feat}}=0.55$) enforces
alignment with channel-blind teacher features, penalizing any student
mechanism that introduces SNR-dependent feature variation. We do not
claim this rules out SNR conditioning in general, only that RCC, which
provides robustness via power asymmetry rather than SNR-dependent
feature shifts, fits this setup better.

\subsection{Rate Analysis}
\label{sec:rate}

Table~\ref{tab:rate} investigates the base-to-residual split at fixed
$d_b+d_r=16$. The base-heavy $(12,4)$ split achieves the highest
BLEU-1 ($0.794$), with the balanced $(8,8)$ split close behind
($0.781$); both outperform the residual-heavy $(4,12)$ configuration
($0.741$). This is consistent with the UEP intuition: the base stream
encodes the semantic core that the residual cannot recover, so
allocating more dimensions to the base matters more than refining an
already-adequate coarse estimate. The $65.5/34.5$ split learned by the
adaptive gate (Sec.~\ref{sec:film}) further supports this hierarchy.
We retain $(8,8)$ as the main configuration because equal stream
capacity makes the decoder-mode ablation a clean attribution test for
UEP and the loss design alone; the small $+0.013$ BLEU-1 gain of
$(12,4)$ is single-seed and is treated as a complementary indicator
rather than the headline configuration.

\begin{table}[!t]
\centering
\caption{Rate analysis: base/residual dimension split (AWGN, 40 epochs).}
\label{tab:rate}
\setlength{\tabcolsep}{4pt}
\begin{tabular}{@{}cccc@{}}
\toprule
$(d_b,\,d_r)$ & \textbf{BLEU-1} & \textbf{BLEU-4} & \textbf{SBERT}\\
\midrule
(4,\,12) & 0.741 & 0.512 & 0.682 \\
(8,\,8) & 0.781 & 0.558 & 0.712 \\
(12,\,4) & \textbf{0.794} & \textbf{0.579} & \textbf{0.723}\\
\bottomrule
\end{tabular}
\end{table}

\subsection{Efficiency and Quality Trade-off}
\label{sec:eff}

We frame the contribution as an efficiency and quality trade-off: the
student preserves semantic quality under channel noise while
reducing inference cost. The realistic deployment metrics are the
single-teacher comparison, since a five-teacher ensemble is not a
practical edge baseline. On an A100 GPU at batch~$32$ decoding $30$
tokens (Table~\ref{tab:compress}), the student takes $86.0$ milliseconds (ms) per forward pass vs.\ $154.0$\,ms
for a single teacher (a $1.79\times$ speedup), and requires $111.3\times10^{6}$
floating-point operations (FLOPs) (computed via \texttt{fvcore}) vs.\ $150\times10^{6}$ for a single
teacher. Halving the Transformer layer count ($4\!\to\!2$) and the
feed-forward width ($512\!\to\!256$) yields $1.33\times$ non-embedding
parameter compression ($3.67\times10^{6}$ vs.\ $4.89\times10^{6}$). The residual codec
keeps $d_b+d_r=16$, matching the teacher channel-symbol dimension at
no additional spectral cost. For completeness, the cost of the
five-teacher ensemble is $777.4$\,ms / $749.4\times10^{6}$ FLOPs ($9.0\times$
and $6.7\times$ the student, respectively).

\begin{table}[!t]
\centering
\caption{Compression and efficiency. Single-teacher and five-teacher-ensemble costs are listed separately; the ensemble is an upper-bound reference, not a deployment baseline. Latency is
 measured on an A100 GPU at batch size $32$ for $30$-token decoding.}
\label{tab:compress}
\setlength{\tabcolsep}{3pt}
\begin{tabular}{@{}lcccc@{}}
\toprule
 & \textbf{1 Teacher} & \textbf{5T Ens.} & \textbf{Student} & \textbf{vs.\ 1T}\\
\midrule
Transformer layers & 4 & 4 & 2 & $2.0\times$\\
Feed-forward dim & 512 & 512 & 256 & $2.0\times$\\
Channel bottleneck & 16 (single) & 16 & 8+8 (RCC) & $1.0\times$\\
Non-embed.\ params & $4.89\!\times\!10^{6}$ & -- & $3.67\!\times\!10^{6}$ & $1.33\times$\\
Total params & $10.6\!\times\!10^{6}$ & -- & $9.36\!\times\!10^{6}$ & $1.13\times$\\
FLOPs (fwd, batch~1) & $150\!\times\!10^{6}$ & $749\!\times\!10^{6}$ & $111\!\times\!10^{6}$ & $1.35\times$\\
Latency (ms, batch~32) & $154.0$ & $777.4$ & $86.0$ & $1.79\times$\\
\bottomrule
\multicolumn{5}{@{}p{0.96\columnwidth}@{}}{\footnotesize The final column is the deployment-relevant single-teacher ratio; against the five-teacher ensemble the student is $6.7\times$ cheaper in FLOPs and $9.0\times$ faster.}
\end{tabular}
\end{table}

\subsection{Limitations}
\label{sec:limitations}

Four limitations are worth noting. First, we assume perfect CSI and
one-tap equalization under flat fading; imperfect-CSI behavior is left
as future work. Second, BLEU-1 can overstate semantic quality;
 Table~\ref{tab:qual} and SBERT serve as complementary evidence, but a
human-rated study is not included. Third, teacher diversity arises
only from random-seed re-initialization. Fourth, the centrality
criterion converges to a single anchor per run, so the method is
empirically a centrality selector rather than a per-batch robust
aggregator.

\section{Conclusion}
\label{sec:conclusion}

We studied a compressed DeepSC student trained by multi-teacher
knowledge distillation, combining a residual channel bottleneck under
unequal error protection with a Krum-inspired centrality-based teacher
selector. On EuroParl, a two-layer student recovers about $98\%$, $93\%$, and $96\%$ of teacher BLEU-1, BLEU-4, and SBERT under AWGN, and about $89\%$, $77\%$, and $87\%$ under Rayleigh, while reducing non-embedding parameters
by $1.33\times$ and single-teacher inference latency by $1.79\times$
($9.0\times$ vs.\ a five-teacher ensemble used as upper-bound
reference). Three-seed reruns indicate complementary gains of $+8.4\%$
BLEU-1 from the residual bottleneck and $+3.0\%$ from centrality-based
selection over mean aggregation, with corroborating SBERT trends. A
decoder-mode ablation isolates the residual stream as a refinement on
top of the base, and a teacher-exclusion experiment shows the selector
tracks the centroid of whichever ensemble is provided. Natural
extensions include adaptive UEP conditioned on instantaneous CSI,
evaluation under imperfect CSI, and the same recipe applied to image
and multimodal semantic communication.

\footnotesize
\setlength{\itemsep}{-2pt}

\end{document}